\documentclass[aps,pra,twocolumn,groupedaddress]{revtex4-1}
\usepackage{setspace}                       
\usepackage{titlesec}   
\usepackage[toc,page,titletoc]{appendix}    
\usepackage{amsmath,amssymb,amsfonts}       
\usepackage{bbold}
\usepackage{dsfont}
\usepackage{graphicx}                       
\usepackage{color}                          
\usepackage{hyperref}                       
\usepackage{calrsfs}
\usepackage{epstopdf}
\usepackage{titlesec}
\usepackage{placeins}
\usepackage{float}

\begin{document}

\title{Affecting Non-Markovian behaviour by changing bath structures}
\author{V. Venkataraman}
\affiliation{QOLS, Blackett Laboratory, Imperial College London, London, SW7 2BW, United Kingdom}

\author{A.D.K Plato}
\affiliation{QOLS, Blackett Laboratory, Imperial College London, London, SW7 2BW, United Kingdom}

\author{Tommaso Tufarelli}
\affiliation{QOLS, Blackett Laboratory, Imperial College London, London, SW7 2BW, United Kingdom}

\author{M.S. Kim}
\affiliation{QOLS, Blackett Laboratory, Imperial College London, London, SW7 2BW, United Kingdom}

\date{\today}

\begin{abstract}
For many open quantum systems, a master equation approach employing the Markov approximation cannot reliably describe the dynamical behaviour. This is the case, for example, in a number of solid state or biological systems, and it has motivated a line of research aimed at quantifying the amount of non-Markovian behaviour in a given model. Within this framework, we investigate the dynamics of a quantum harmonic oscillator linearly coupled to a bosonic bath. We focus on Gaussian states, which are suitably treated using a covariance matrix approach. Concentrating on an entanglement based non-Markovian behaviour quantifier (NMBQ) proposed by Rivas et. al. \cite{NMmeasure}, we consider the role that near resonant and off-resonant modes play in affecting the NMBQ. By using a large but finite bath of oscillators for both Ohmic and super Ohmic spectral densities we find, by systematically increasing the coupling strength, initially the near resonant modes provide the most significant non-Markovian effects, while after a certain threshold of coupling strength the off-resonant modes play the dominant role. We also consider the NMBQ for two other models where we add a single strongly coupled oscillator to the model in extra bath mode and `buffer'  configurations, which affects the modes that determine non-Markovian behaviour. 
\end{abstract}

\maketitle

\titlespacing*{\section}{0pt}{10pt}{10pt}
\section{Introduction}
Many realistic quantum-mechanical models are formulated in the framework of open quantum systems. Indeed, it is a significant experimental challenge to isolate the quantum systems of interest from their environment. A standard way to describe this type of dynamic is the master equation approach, where the environmental modes are traced out to leave an equation of motion for the density matrix of the system. In order to make the problem tractable, this is typically combined with a series of approximations, one of which is the Markov approximation. This implies that the evolution of the system is only dependent on its current state, i.e. the future dynamics does not depend on its previous trajectory. A typical formalisation of this notion is the definition of Markovianity can be extended to signify that the system evolution is `divisible', a term that we shall specify below.
The Markov assumption is not valid for some models, e.g. dealing with biological or solid-state systems \cite{bio1,bio2,bio3,ss1,ss2}, where the effect of non-Markovian behaviour (NMB) of the environment cannot be neglected. Among other reasons, this has generated a significant amount of literature dealing with the quantification of the degree of NMB present in an open system with the first non-Markovian (NM) measure developed for Gaussian channels by Wolf et. al. \cite{Wolf}, followed by further proposals which utilise quantities such as entanglement, trace distance, fidelity and Fisher information \cite{NMmeasure,breuer1,breuer2,Fisher}. 

The motivation for this work is to understand the effect various bath modes have on NMB. By changing the coupling of these modes and the structures of the bath we can significantly affect the NMB of the model. Not only is this exercise useful for our understanding of NMB but it could be utilised to manipulate how and when quantum information (particularly entanglement) can be fed back to the system. In this paper we apply these ideas to the paradigmatic scenario of a quantum harmonic oscillator coupled to a bath of bosonic modes, with beam-splitter like interactions.  We restrict our attention to Gaussian states, which allows us to handle a large but finite number of oscillators via the covariance matrix formalism. By considering baths characterized by either Ohmic or super Ohmic spectral densities we aim to understand the role that near resonant and off-resonant modes in the environment play in determining the degree of NMB predicted by one of the aforementioned `measures'. At a mid range coupling strength, we observe that the near resonant bath modes provide the largest contribution to the NM character of the dynamics. On the other hand, as the system-environment coupling strength is increased, a larger number of the off-resonant modes take active part into the system's evolution, superseding the near resonant modes in determining the NM character of the system. 

We begin with the definition of non-Markovianity in order to quantify the NMB in our models. Our analysis starts with the entanglement dynamics of systems which consider only a pair of coupled harmonic oscillators. This will give us the intuition to determine the factors that affect the entanglement-based NM `measure' for models with many modes. In this paper we investigate four models, the first two are the common cases of an oscillator coupled a large finite bosonic bath, for Ohmic and super Ohmic spectral densities. The other two cases add a strongly coupled resonant mode to the system in a way which affects the NM `measure' in an interesting fashion. Throughout this paper we will use dimensionless units for all our parameters such as frequency and coupling strength. 
\titlespacing*{\section}{0pt}{10pt}{10pt}
\section{Definition and quantification of non-Markovianity}
To gain a {\it quantifier} for the degree of NMB in an open 

\noindent quantum system we shall make use of a sufficient condition based on entanglement \cite{NMmeasure} (and fidelity \cite{breuer2} in Appendix E). Before illustrating the quantifier in detail, let us concentrate on the definition of Markovianity that we adopt throughout the paper, which is the one used in Ref. \cite{NMmeasure}. Note that there are measures which utilise other definitons, for example based on information back-flow \cite{breuer2, breuer1}. 

The dynamics of a quantum system is described in general via a completely positive, trace-preserving map (CPT) $\mathcal{E}_{(t_f,t_i)}$, such that if a state $\rho_i$ is prepared at an initial time $t_i$, the corresponding state at a later time $t_f$ is given by $\rho_f\equiv\mathcal{E}_{(t_f,t_i)}(\rho_i)$. If we fix a start time $t_0$, and a final time $t_2$ the map $\mathcal E$ is {\it Markovian between $t_0$ and $t_2$} iff, for any $t_1$ in the interval $t_0\leq t_1\leq t_2$, the following composition law holds
\begin{equation}\label{complaw}
\mathcal{E}_{(t_{2},t_{0})}=\mathcal{E}_{(t_{2},t_{1})}\circ\mathcal{E}_{(t_{1},t_{0})}.
\end{equation}
where $\mathcal{E}_{(t_{1},t_{0})}$ and $\mathcal{E}_{(t_{2},t_{1})}$ are CPT. This divisibility property attempts to formalise the memoryless interpretation of Markovianity. Indeed, the implications of Eq.~\eqref{complaw} can be understood via the following simple example. Consider an initial state $\rho_0$ and define $\rho_1\equiv \mathcal{E}_{(t_{1},t_{0})}(\rho_0)$ and $\rho_2\equiv\mathcal{E}_{(t_{2},t_{0})}(\rho_0)$. Eq.~\eqref{complaw} would suggest that there exists a CPT map $\mathcal{E}_{(t_{2},t_{1})}$ that takes the state $\rho_1$ to a state $\rho_2$ without knowledge of the history prior to $t_1$. This illustrates that it does not matter {\it how} the system has evolved between $t_0$ and $t_1$, and only the knowledge of the system at time $t_1$ is required to determine its evolution between $t_1$ and $t_2$.

From now on, we take the violation of Eq.~\eqref{complaw} as our definition of non-Markovianity. Let us now introduce the NM quantifier we employ for our investigations. We shall consider the entanglement based NM sufficient measure proposed by Rivas et. al. \cite{NMmeasure}. They consider a bipartite system comprising of the system under investigation plus an ancilla. The two are initially prepared in a two-mode squeezed state $\rho_{\sf SA}(0)$, and their entanglement is tracked as a function of time. Keeping in mind that no local operation and classical communication (LOCC) operation can increase entanglement \cite{entangmeasures}, any system evolution satisfying the divisibility property \eqref{complaw} dictates that a system-ancilla entanglement would monotonically decrease with time.
If instead an increase in entanglement is detected, Eq.~\eqref{complaw} must necessarily be violated, ergo the dynamics has to be NM.  

 Choosing an appropriate entanglement measure $E$, one may quantify NMB by summing up all the entanglement increases detected during the time interval
of interest. Hence, the NM quantifier is defined as
 \begin{equation}\label{NMmeasure}
    \mathcal{I}^{(E)}\equiv E[\rho_{\sf SA}(t_{\sf f})]\!-\!E[\rho_{\sf SA}(t_{0})]+\!\int_{t_{0}}^{t_{\sf f}}\!\Big\vert\frac{dE[\rho_{\sf SA}(t)]}{dt}\Big\vert dt.
 \end{equation}
Throughout this paper we use logarithmic negativity as the entanglement measure since it is easily computable for a Gaussian state \cite{Logneg1, Logneg2, covareq}.

It is important to note at this point that $\mathcal{I}^{(E)}>0$ defines only a {\it sufficient} condition for NM, hence the above quantity should be interpreted as an NMB {\it quantifier} (NMBQ), rather than a full-fledged measure,  which can be gained if the full dynamical map is known (as shown in Ref. \cite{NMmeasure} with the use of the Choi - Jamiolkowski isomorphism). We chose to use this quantifier because it was easy to calculate and we can directly understand how NMB could be utilised to control the flow of entanglement, a useful resource in quantum information. 

\titlespacing*{\section}{0pt}{10pt}{10pt}
\section{Analysis of coupled oscillators}
\begin{figure}[h!]
\centering
\includegraphics[scale=0.4]{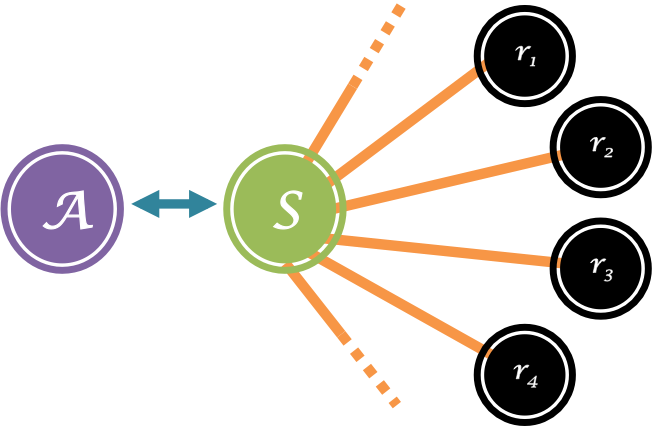}
\caption{Diagram of Model 1. An ancilla (A) is entangled with the system (S) (represented by the double arrow). Orange lines indicate coupling between the system and each bath mode ($r_{i}$).} \label{Model1}
\end{figure}
Let us consider a two-mode (ancilla (A) and system (S)) squeezed state defined as $S_{AS}(\zeta)\vert00\rangle_{AS}$ \cite{BarnettRadmore} where $S_{AS}(\zeta)$ is the two-mode squeezing operator with the squeezing parameter $\zeta$. It is well-known that the two-mode squeezed state is entangled for any $\zeta \neq 0$ \cite{rsimon}. If we assume that the system mode, of this squeezed state, interacts with a bosonic bath and the ancilla is left intact, we are left with a Hamiltonian of the form,
\begin{equation}\label{genham}
     H = \omega_{a}a^{\dagger}a + \omega_{s}s^{\dagger}s + \sum_{i=1}^{N}\omega_{r_{i}}r_{i}^{\dagger}r_{i} + \sum_{i=1}^{N}g_{i}(s^{\dagger}r_{i} + s r_{i}^{\dagger}).
\end{equation}
where {\it $g_{i}$} is the system-bath mode coupling strength and $a$, $s$ and $r_{i}$ (and their adjoints) are the annihilation (and creation) operators for the ancilla, system and bath modes respectively. Throughout the paper we fix the system and ancilla frequencies ($\omega_{s}$ and $\omega_{a}$) to 10 in all the models. Note that we have taken a rotating wave approximation in the interaction. This common model (Fig. \ref{Model1}) will allow us to use the NMBQ to witness the change in NMB as we vary bath parameters and structures. But as our bath is comprised of many modes interacting with the system, it would be difficult to discern how the different modes in the bath can affect the NMBQ individually. To this end, we initially begin with a `bath' of just one oscillator. This toy model does not reflect in any way the many mode nature of the bath, but we can gain an insight into the dynamics that could be at play in an open quantum system, which we consider later. 

\titlespacing*{\subsection}{0pt}{10pt}{10pt}
\subsection{Single oscillator}

\begin{figure}[h!]
\centering
\begin{center}
\includegraphics[width=0.45\textwidth,height=4.5cm]{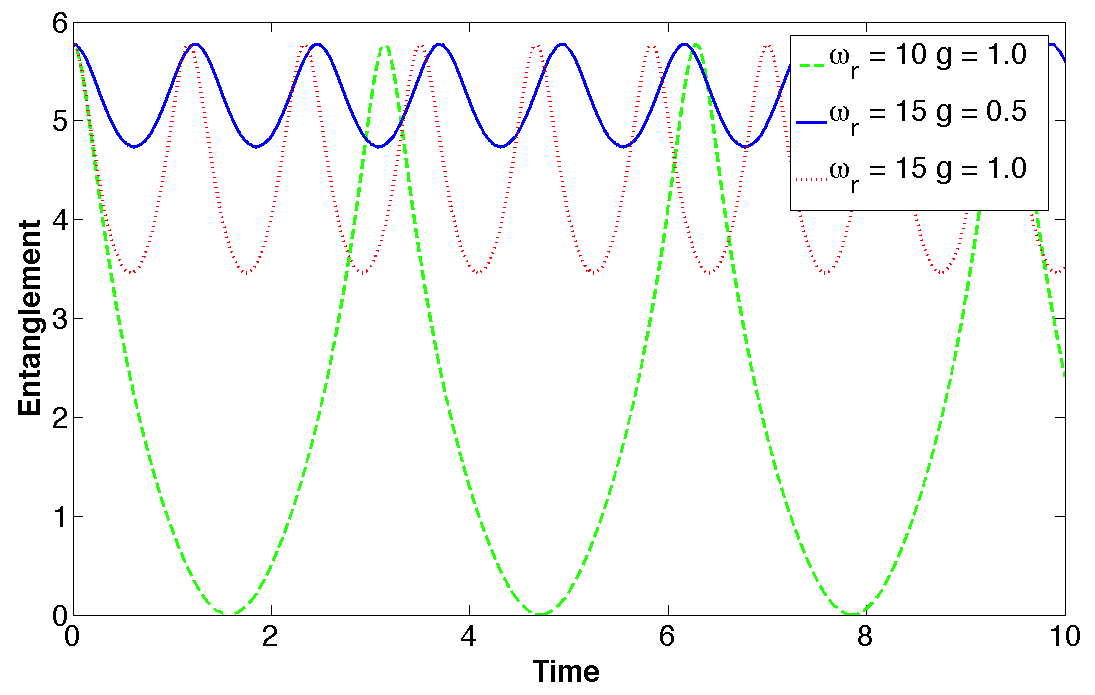}
\end{center}
\caption{Entanglement dynamics for a system-ancilla state, with $\omega_{s}=10$, coupled to an oscillator (r) with the properties; Green - [$\omega_{r}= 10$ , $g = 1$], Blue - [$\omega_{r}= 15$ , $g = 0.5$], Red - [$\omega_{r}= 15$ , $g = 1$]. The squeezing parameter $\zeta = 4$. The single mode (r) is an a thermal state with temperature, T=1.}\label{2modent}
\end{figure}

In a single oscillator model, i.e. Eq.~\eqref{genham} with N = 1, the entanglement will oscillate between the system and the single mode (r) during the time evolution. Note that the ancilla is not coupled to any oscillator and only undergoes free evolution. The ancilla-system construct is a tool to witness the non-divisibility of the system's dynamics. The frequency of the ancilla-system entanglement oscillation (EO) is representative of the speed at which information travels between the system and the oscillator. If {\it g} is increased then the information travels back and forth faster, i.e. an increase in EO frequency. We also have to consider $\omega_{r}$ ($\omega_{r_{1}}$) in relation to $\omega_{s}$. For resonant interactions ($\omega_{r} = \omega_{s}$) we find that the entanglement is shared maximally regardless of {\it g}. If the oscillator is detuned ($\omega_{r} \neq \omega_{s}$) we find that the EO increases in frequency but the magnitude of the EO decreases, seen by comparing green and red lines in Figure \ref{2modent}. For detuned modes, increasing the coupling strength also increases the magnitude of the EO and this behaviour is clearly shown by the blue and red lines in Figure \ref{2modent}.
 For the case of large detuning ($\omega_{r}-\omega_{s} = \Delta \gg g$) an analytic expression for the EO can be found (Appendix B),
\begin{equation}
\mathbb{E}= \log_{2}\left(\frac{\Delta^{2}e^{-\zeta}+ 2g^{2}\left(\coth\left(\frac{w_{r}}{2}\right) - e^{-\zeta} \right)\sin\left(\frac{\Delta t}{2}\right)}{2\Delta^{2}}\right).
\end{equation}
From the single oscillator model we can gain an insight into the predictions of the NMBQ for a many oscillator bath. It is important to note that we can only gain an intuition for the dynamics at play since adding even a single oscillator to the `bath' complicates the dynamics significantly, as we have shown analytically in the Appendix C.  
The NMBQ sums up all entanglement increases and will therefore depend on two aspects of the EO, the magnitude and the frequency. At a low coupling strength a near resonant mode would yield more NMB than an off-resonant mode. This is due to the much larger EO magnitude of a near resonant mode. But as {\it g} is pushed past a specific value for a particular detuning, the detuned mode would yield more NMB due to a combination of the high frequency and increased magnitude of the EO. Keeping this in mind, we now investigate the behaviour when there are many oscillators coupled to the system in order to see how the NMBQ is affected.  
 
\titlespacing*{\subsection}{0pt}{10pt}{10pt}
\subsection{Many bath mode models} 
\begin{figure}[h!]
\centering
\includegraphics[scale=0.4]{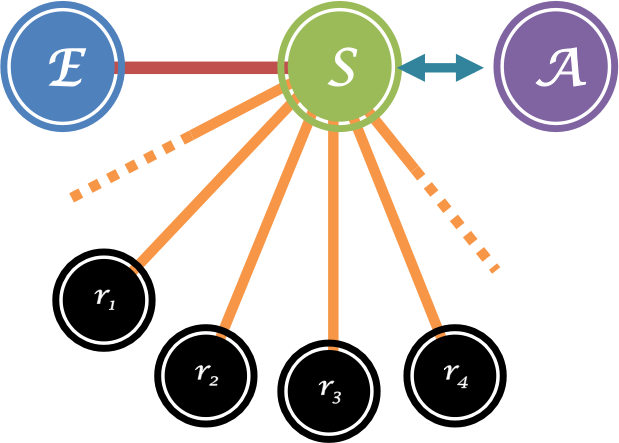}
\caption{Diagram of Model 2. Model 2 is similar to Model 1, with a system (S) - ancilla (A) entangled state and the system coupled to all bath modes ($r_{i}$), but now there is an extra resonant mode (E) in the bath with a fixed coupling strength of 1.} \label{Model2}
\end{figure}

\begin{figure}[h!]
\centering
\includegraphics[scale=0.4]{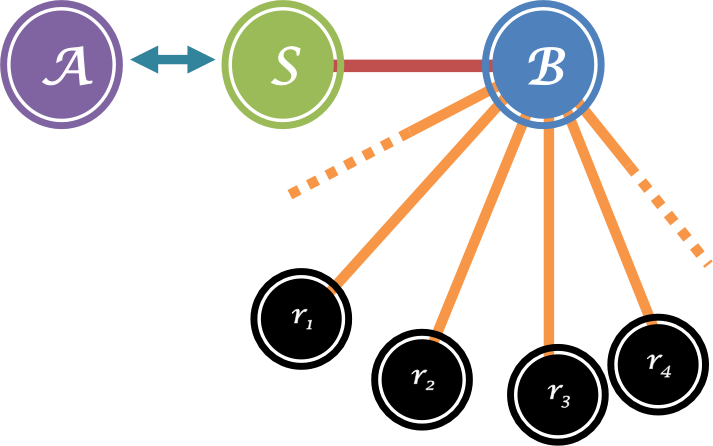}
\caption{Diagram of Model 3. This model consists of the same parts as Model 2 but in a different configuration. The resonant extra mode is now a resonant buffer (B) which is coupled to every bath mode ($r_{i}$) as well as the system (S). The buffer-system coupling strength is set to 1. } \label{Model3}
\end{figure}

In this section we will consider four different types of interaction between the bath modes and the system. These different models will show how bath structures can be manipulated in order to affect the NMBQ. Model 1 (Figure \ref{Model1}) shows the role of near resonant and off resonant modes in the bath. Models 2 and 3 (Figures \ref{Model2} and \ref{Model3}) displays how adding a single strongly coupled resonant mode can affect the NMB. 

\begin{figure}
\centering
\begin{center}
\includegraphics[width=0.45\textwidth,height=4cm]{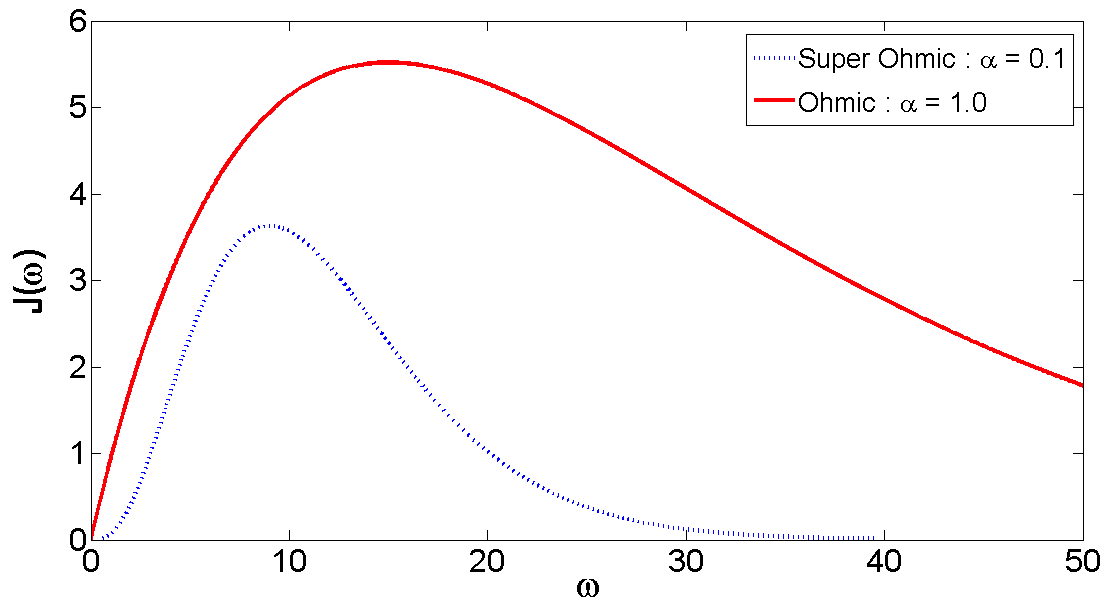}
\end{center}
\caption{Form of the spectral density functions for the Ohmic and super Ohmic cases with cut-off frequencies, $\omega_{c}$, 15 and 3 respectively.}\label{SpecDen} 
\end{figure}

The first two types of interaction we consider are for Model 1. In the Hamiltonian given in Eq. (\ref{genham}) the coupling strengths of the bath modes $g_{i}$ are determined by the spectral density function of the bath $J(\omega)$ \cite{Dougpaper}, 
\begin{equation}
    g_{i}^{2}\approx J(\omega_{r_{i}}) \Delta\omega.
\end{equation}
We consider both Ohmic (Eq. (\ref{ohmic})) and super Ohmic (Eq. (\ref{superohmic})) spectral densities, with an exponential cut-off , $\omega_{c}$ (ensuring that the high frequency couplings do not diverge), and a damping factor, $\alpha$. 
\begin{equation}\label{ohmic}
J(\omega_{r_{i}})_{\it{O}}=\alpha\omega_{r_{i}} e^{-\omega_{r_{i}}/\omega_{c}}
\end{equation}
\begin{equation}\label{superohmic}
J(\omega_{r_{i}})_{\it{SO}}=\alpha\omega^{3}_{r_{i}} e^{-\omega_{r_{i}}/\omega_{c}}
\end{equation}
The forms of these functions are shown in Figure \ref{SpecDen}. The cut-off frequencies, $\omega_{c}$, for the Ohmic and super Ohmic baths are 15 and 3 respectively. The other two cases we consider are for Models 2 and 3 with an Ohmic spectral density.

To evaluate the NMBQ, we simulate the system coupled to a bath of 350 oscillators, in contrast to a master equation formalism to avoid using approximations. The frequencies of the bath oscillators are distributed evenly up to a maximum frequency $\omega_{bmax}$, the frequency splitting $\Delta\omega$ is therefore given by $350/\omega_{bmax}$. The initial state of the bath for all the models is a thermal state, with temperature $T$, and the ancilla-system is a two mode squeezed state. In models 2 and 3, the additional bath and buffer modes are also initially thermal with temperature $T$. We fix the system and ancilla frequencies, $\omega_{s} = \omega_{a} = 10$, the maximum bath mode frequency, $\omega_{bmax}= 50$, the temperature, $T=1$ and the squeezing parameter $\zeta=4$ for Models 1, 2 and 3. Numerical results indicate that the squeezing parameter acts only to rescale the NMB without losing the qualitative features and so we chose $\zeta$ to exaggerate the observed effects (though not so high as to cause problems with the numerics). The details of the simulation can be found in Appendix A. Figure \ref{NMallmodels} shows the predictions of the NMBQ for the three Ohmic bath models as a function of the spectral density damping factor $\alpha$. 
\begin{figure}[h!]
\centering
\includegraphics[width=0.45\textwidth]{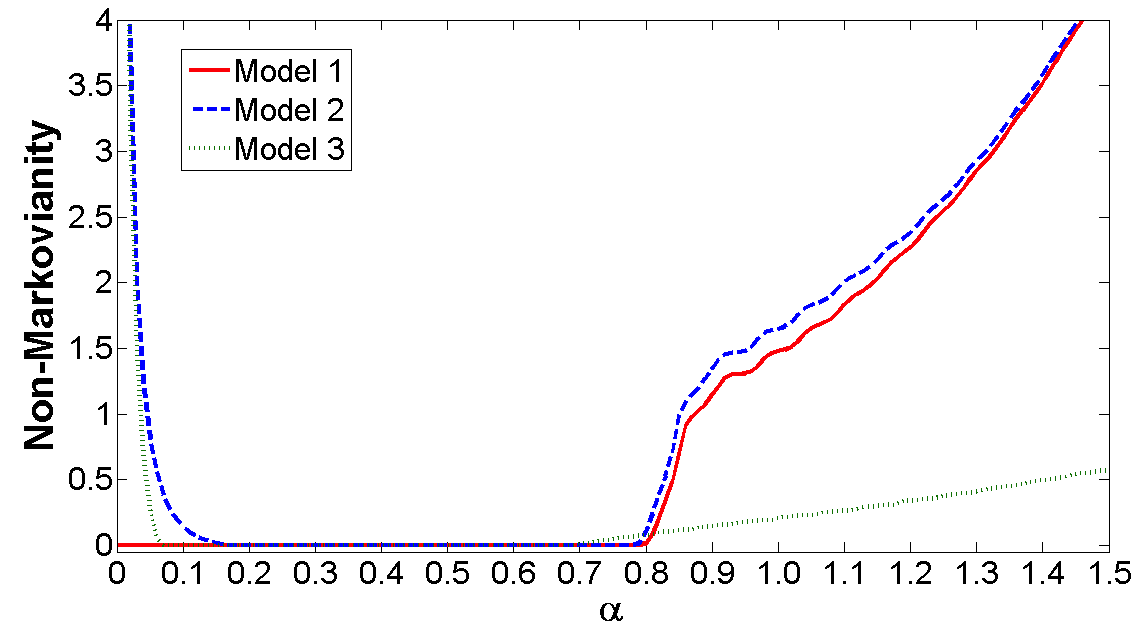}
\caption{The figure shows the NMBQ for the three models. We see that there is a high threshold $\alpha$ value for NMB and Models 2 and 3 display NMB for low $\alpha$ due to the strong coupling to the additional mode. The simulation is run from time $t_{0}=0$ to $t_{f}=20$ in time intervals of $\Delta t = 0.001$. }
\label{NMallmodels}
\end{figure}

\begin{figure}[h!]
\centering
\includegraphics[width=0.45\textwidth]{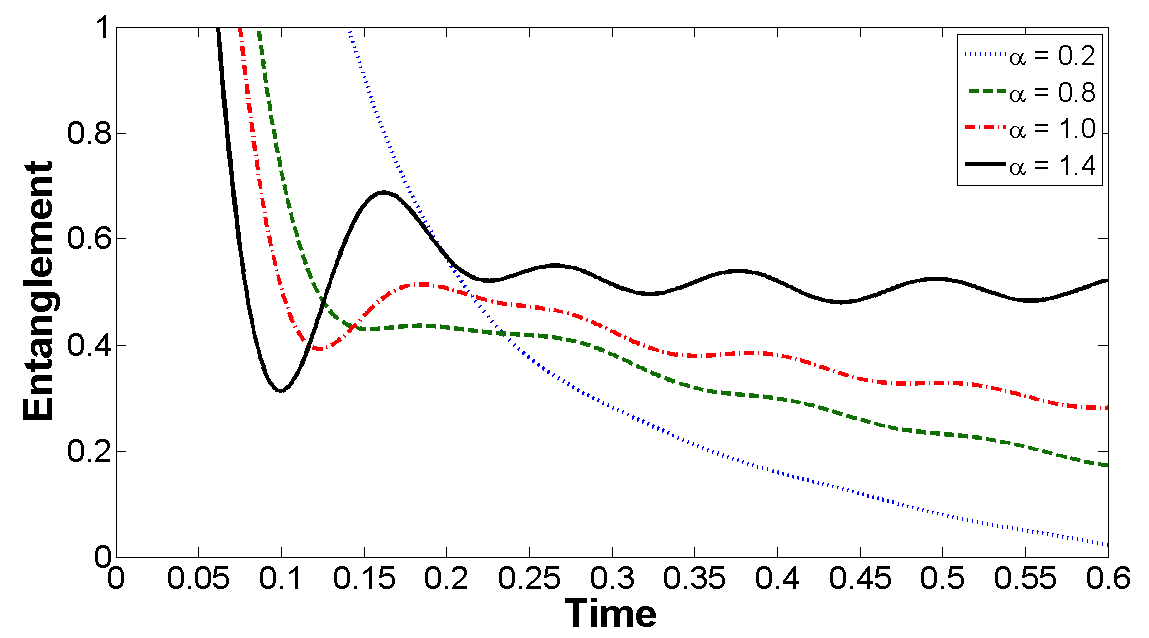}
\caption{Entanglement dynamics for Model 1 with an Ohmic bath. Coupling strengths to bath modes are varied by changing $\alpha$ in the spectral density function. We detect NMB after an $\alpha$ value of 0.8 and significant oscillations in the entanglement are seen after this value.} \label{EntOsc}
\end{figure}

The figure shows distinct regions of NMB. To understand these different regions for the models we can consider the entanglement dynamics for varying values of $\alpha$. Using our knowledge of coupling strengths and the occupation numbers of the bath modes with time (see Appendix D), we can construct an interpretation of the processes involved when there are numerous modes coupled to the system. Starting with Model 1, Figure \ref{EntOsc} shows the entanglement dynamics for the Ohmic case for varying $\alpha$. 
 
As time passes the entanglement is shared, unequally, to all the modes in the bath and sometimes this entanglement comes back to the system (if at all). The result of this dynamic depends on various parameters, which consequently decide the NMB of the model. 

In the Ohmic case, when the bath is weakly coupled with a scaling factor of $\alpha=0.2$, the spectral density function would suggest that the near resonant bath modes have the strongest coupling and therefore the highest occupation and indeed we see that in Figure \ref{Mod1occ}. At this stage the coupling strengths to the bath modes are too weak and  we see decoherence, which leads to dynamics that do not produce NM effects detectable by the NMBQ. As $\alpha$ is increased, the entanglement starts to decay faster due to a stronger coupling to the bath i.e. a faster transmission of information to the bath where it decoheres. For higher values of $\alpha$ however, EO are increasingly found and NMB is detected by the NMBQ.

The non-Markovianity of the model in this region of $\alpha$ values is influenced by a variety of factors, including the strength of system-bath couplings, the occupation numbers of the bath modes, and the ability of the system to induce oscillations in a bath mode's occupancy (which is an indicator of the level of interaction between them). If $\alpha$ is increased beyond a certain threshold, initially a situation arises where the profile of the spectral density dictates that the system is {\it significantly} strongly coupled to near resonant modes. Since the system shares more entanglement with the near resonant modes, the stronger coupling increases the likelihood that the dynamic results in an entanglement increase for the system-ancilla state. For now it is the near resonant modes that are the main contributors to the NMB. This is due to the fact that at this coupling strength the combined frequency and magnitude of their EO is more than that of the detuned modes. Because the detuned modes have a very low EO magnitude thanks to the relatively weaker coupling they are allocated by the Ohmic function. 

As dictated by Eq. \ref{ohmic}, when $\alpha$ is increased further, the detuned modes begin to couple more strongly, resulting in increased magnitude of their EO. Our intuition is that if we include the fact that they have high frequency EO (as a result of the detuning) and that they greatly outnumber the near resonant modes, the combined entanglement increases of the detuned modes will be greater than the near resonant modes. This now makes the detuned modes the important players in determining the NMB of the model. The importance of detuned modes at stronger couplings is shown in Figure \ref{Mod1occ}, where it can be seen that an increase in $\alpha$ results in more occupancy and increased occupancy oscillations of these modes. This indicates that the system is interacting more with these modes. Moreover we notice at certain times the near resonant modes are not occupied when we see an EO, e.g. at $\alpha = 1$ between times $0.45$ and $0.5$. 

This transition between the importance of near resonant/detuned modes is not easily seen for the Ohmic case, because the occupancy would suggest that it is only the detuned modes which are important. But as we will see in the other models, near resonant modes do have a role to play.

\begin{figure}[h!]
\centering
\includegraphics[width=0.45\textwidth]{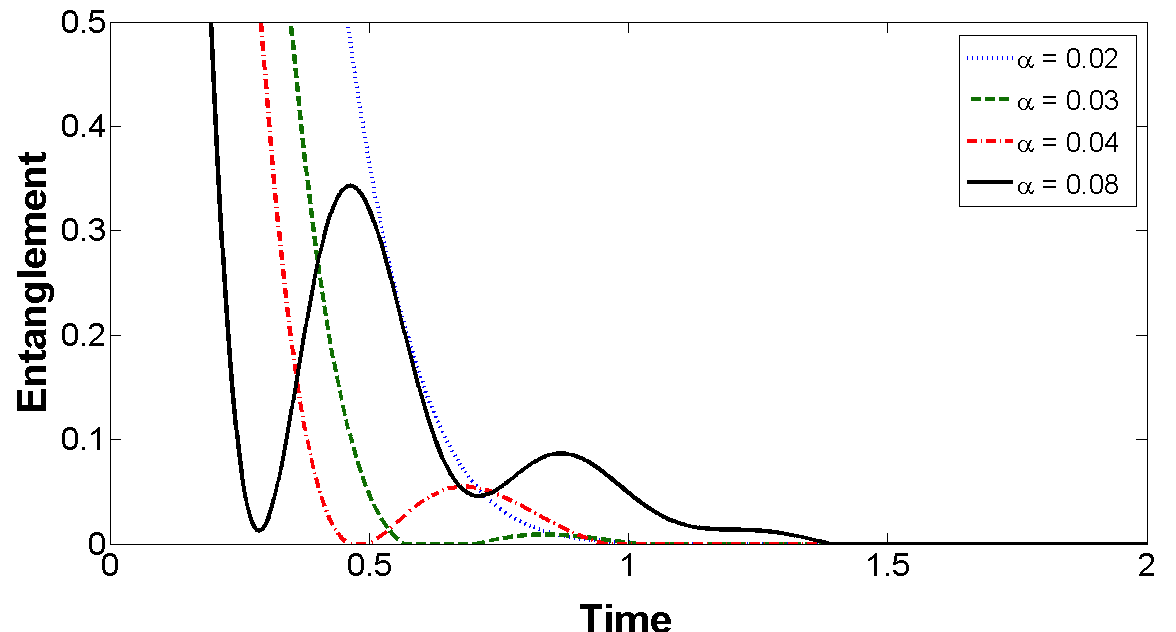}
\caption{Entaglement dynamics for a super Ohmic spectrum. NMB is detected after an $\alpha$ value of 0.03 and as expected we see EO for those values.}
\label{EntOscSO}
\end{figure}

For Model 1 with a super Ohmic spectral density, smaller values of $\alpha$ are needed for strong coupling strengths (see Eq. (\ref{superohmic})). Therefore, as Figure \ref{EntOscSO} shows, a lower threshold $\alpha$ value was needed to observe NMB. Unlike the previous case, Figure \ref{Mod1SOocc} shows that there is still a significant occupation in the near resonant region where we initially witness NMB and due to their naturally large EO magnitude, they play an important role. But as we saw in the Ohmic case, as $\alpha$ is increased further the occupancy and occupancy oscillations of the detuned modes (Figure \ref{Mod1SOocc}) become more significant and they will take the lead.  

\begin{figure}[h!]
\centering
\includegraphics[width=0.45\textwidth]{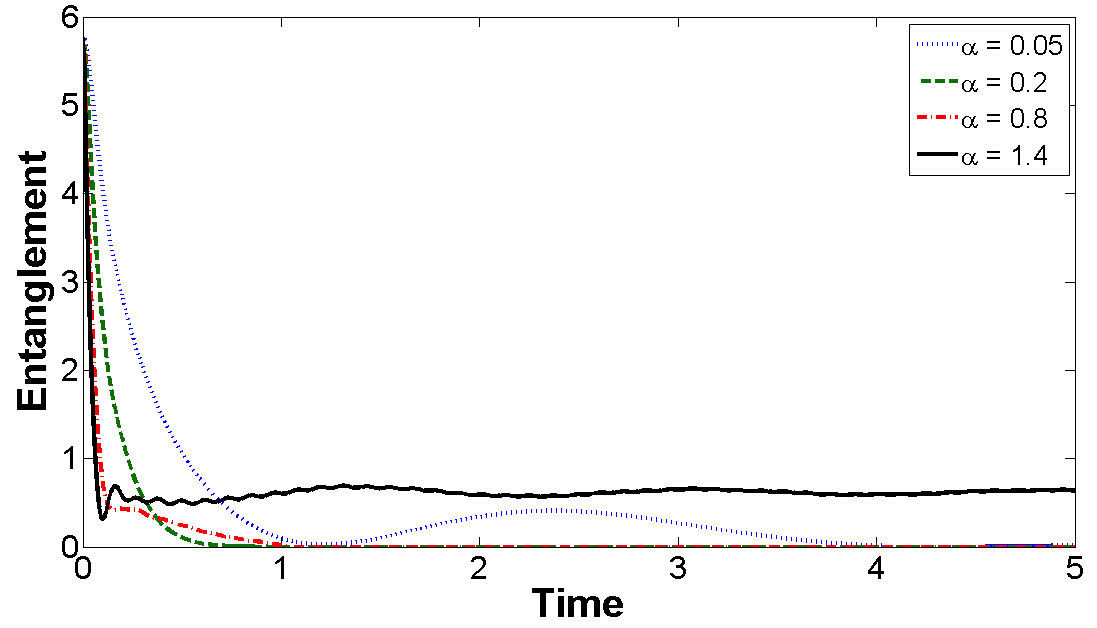}
\caption{Entanglement dynamics for Model 2. It shows that for very low coupling we get EO from the extra mode and as the coupling is increased this oscillation is suppressed. Then as $\alpha$ is increased further we see similar oscillations to the Ohmic case in Model 1.}
\label{Mod2ent}
\end{figure}

If we now consider Model 2, the NMBQ suggests two regions of NMB which can be seen in Figure \ref{NMallmodels}. The first region is for low $\alpha$ values where we get NMB due to the extra strongly coupled resonant mode in the bath. This can be seen from the EO caused by the extra mode in Figure \ref{Mod2ent} and the lack of the occupancy in the resonant mode in the bath (Figure \ref{Mod2occ}), indicating that the extra mode is strongly interacting with the system. As $\alpha$ increases the model behaves like Model 1 where we enter a region where there is no NM dynamics as the rest of the bath is coupled strongly enough to kill the EO from the extra mode. Then, as before, we see that when $\alpha$ is increased beyond a threshold we get NMB according to the same reasoning as in Model 1. Indeed we can see that the NMBQ values follow a similar profile to that of Model 1 but with slightly more NMB. This is shown in the entanglement dynamics (Figure \ref{Mod2ent}) and the occupancy of the bath modes (Figure \ref{Mod2occ}). The additional NMB we notice is due to the extra strongly coupled mode, indicated again by the diminished occupancy in the resonant region compared to that of Model 1. Note that for Model 2 as we reach very high $\alpha$ values the occupancy in the resonant region of the bath increases and therefore Models 1 and 2 have increasingly similar NMBQ values. This is because the resonant extra mode plays a less significant role since the off resonant modes are the greatest contributers to the NMB in the high $\alpha$ region.

\begin{figure}[h!]
\centering
\includegraphics[width=0.45\textwidth]{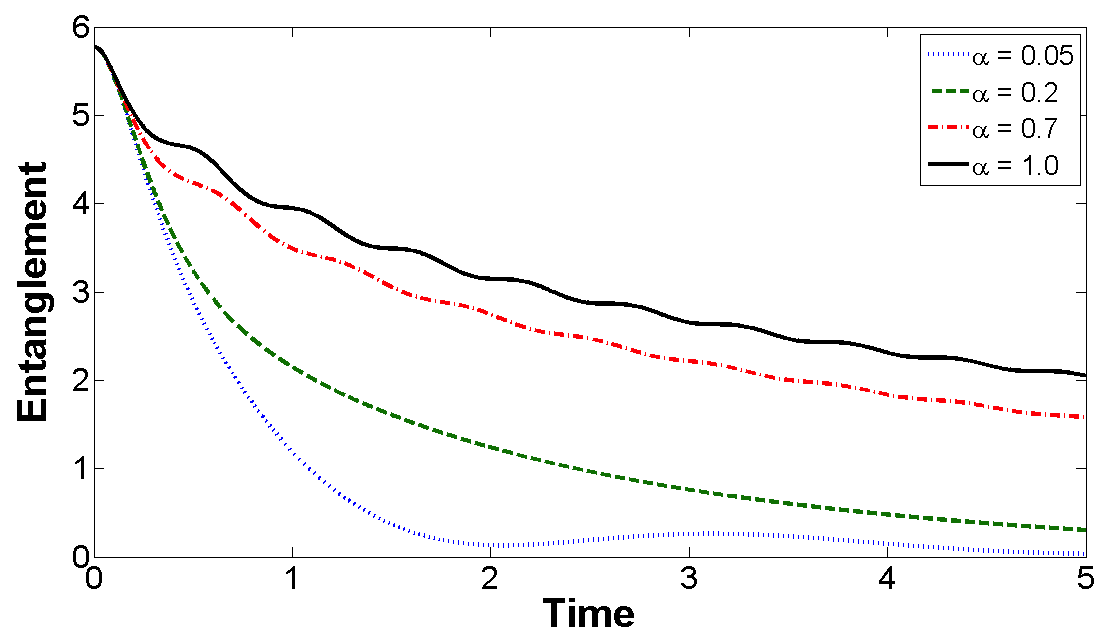}
\caption{Entanglement dynamics for Model 3. It shows that for very low coupling we get EO from the buffer mode and as the coupling is increased this oscillation is suppressed. Then as $\alpha$ is increased further we see oscillations due to the near resonant bath modes.}\label{Mod3ent}
\end{figure}

Model 3 also displayed two regions of NMB as shown in Figure \ref{NMallmodels}. The first is for very low coupling strengths (i.e. low $\alpha$) between the buffer and the bath. We witness NMB due to the `reflections' of the entanglement from the strongly coupled resonant buffer mode (in similar fashion to Model 2), which can be clearly seen in Figure \ref{Mod3ent}. As the buffer-bath coupling strength is increased, NM dynamics is not detected because the buffer leaks the entanglement to the bath before it has a chance to return (i.e. Model 1 at low $\alpha$ and Model 2 in the Markov region). Note however, in comparison to Model 2, a smaller $\alpha$ is needed to see no NMB as the buffer leaks the entanglement to the near resonant modes in the bath before it can give it back to the system mode. 

As we have seen before, beyond a threshold, increasing $\alpha$ results in NMB. Note however that there are two key differences to Models 1 and 2; a lower threshold value of $\alpha$ and a lower value of non-Markovianity. The reasons for these differences can be seen from the occupancy (Figure \ref{Mod3occ}). They show a large occupancy of the near resonant frequency region indicating that the buffer is primarily interacting with the near resonant modes, which are now solely responsible for the NMB. 

Figure \ref{Mod3ent} shows that the EO at high $\alpha$ are significantly different to the other models. We can clearly see that there are fewer oscillations and a longer decay time. This, along with Figure \ref{Mod3occ}, seem to indicate that the buffer has effectively reduced the size of the bath around the resonant region. For the chosen system-buffer coupling, we can hypothesize that the threshold is lower because of the reduced bath size. Also the value of the NMB is lower because the near resonant modes have a lower EO frequency and are few when compared to the detuned modes.

Note that models similar to Model 3 have been investigated in works on effective spectral densities and the structure of the bath which could be used to extend the model \cite{Martinazzo2011, Chin2010}. These papers use techniques that transform the multiple oscillator bath model to a coupled chain. In a different vein of investigation, an analysis of the entanglement dynamics for coupled cavity fields in various baths using Feynman - Vernon influence functional theory \cite{An2007} has been done, as well as EO in a single qubit-bath model \cite{qubitEO}.
\titlespacing*{\section}{0pt}{10pt}{10pt}
\section{Summary}
We investigated the role that near resonant and detuned modes play in determining NMB. The investigation was based on an entanglement based quantifer of non-Markovianity. By considering a two mode model, we noted how the system shares entanglement with modes depending on the strength of the coupling and the level of detuning. Looking at an harmonic oscillator coupled to a bath model, we found that the shape of the spectral density function determined which modes played the important role in NMB. For low $\alpha$ couplings no NMB was observed, but after a threshold value, near resonant modes induce some NMB in the model. However, for higher values of $\alpha$ the detuned modes take over from the near resonant modes and play the lead role in NMB due to the high frequency of their EO. The role of near resonant modes was made more apparent in the other models that we considered. The schemes added a single strongly coupled resonant oscillator which lead to NMB detected at low $\alpha$ due to this mode. As $\alpha$ is increased there is a period where no NM dynamics is detected. But for higher $\alpha$'s, the case which had an extra strongly coupled mode in the bath behaved like the original model but with a increased amount of NMB due to the extra mode. When this extra mode was a buffer between the system and bath, a diminished NMB was noted and was due to the buffer primarily interacting with near resonant modes, which naturally produce less NMB than detuned modes in the strong coupling regime.

Using the knowledge we have gained about the role of near resonant and off-resonant modes in determining NMB, one can isolate the significant modes in a bath containing numerous modes. Armed with these insights we are better equipped to engineer models to control the flow of quantum information in our system. This could be achieved by changing the coupling/detuning to the important modes and/or by adding modes in various configurations. Therefore this type of control could be useful in maintaing the quantum information of the system, for example in quantum memory models, or even to minimise information feedback which can be of assistance in state transfer protocols.
\titlespacing*{\section}{0pt}{10pt}{10pt}
\section{Acknowledgments}
This work has been funded by the EPSRC as part of the Controlled Quantum Dynamics Centre for Doctoral Training (CQD-CDT). The authors would like to thank Ahsan Nazir and Marco Genoni for their time and insight during many fruitful discussions.  MSK and TT would also like to thank the Qatar National Research Fund (NPRP4-554-1-084).
\titlespacing*{\section}{0pt}{10pt}{10pt}
\section{Appendix A - Details of the simulation}
To simulate the models we use a covariance matrix approach which is valid for Gaussian states and Hamiltonians of bi-linear form \cite{Dougpaper, GaussianReview}. The process involves writing the Hamiltonian in the following form
\begin{equation}
    H=\frac{1}{2}R^{T}KR
\end{equation} 
where $K$ is a time independant matrix and $R$ is the vector
\begin{equation}
    R^{T}=(\hat{x}_{a},\hat{x}_{s},\hat{x}_{r_{i}},\hat{p}_{a},\hat{p}_{s},\hat{p}_{r_{i}})
\end{equation}
which contains the position and momentum operators for the ancilla ($a$), system ($s$) and the bath ($r_{i}$) modes. The $R$ vectors obey the commutation
\begin{equation}
    [R_{a},R_{b}]=i\sigma_{ab},
\end{equation}
where $\sigma$ is the symplectic matrix
\begin{equation}
    \sigma=\left(
             \begin{array}{cc}
               0 & \mathbb{1}_{n} \\
               -\mathbb{1}_{n} & 0 \\
             \end{array}
           \right)
\end{equation}
and $n$ is the number of modes in the model. The Heisenberg equation for $R$ can be solved to find the time evolution of the vector
\begin{equation}\label{Revo}
    R(t)=e^{\sigma Kt}R(0).
\end{equation} 
The covarince matrix describes the state of the system \cite{GaussianReview} and can be written in the form
\begin{equation}\label{Cmat}
     \gamma_{jk}=2Re\text{Tr}{[\rho R_{j}R_{k}]}.
\end{equation}
Using equations \ref{Revo} and \ref{Cmat} we can find the time evolution of the covariance matrix
\begin{equation}
\gamma (t) = e^{\sigma K t}\gamma (0) e^{-K\sigma t}
\end{equation}
which we can simulate using MATLAB to find the state of the ancilla-system at any time. For example the $K$ and $\gamma (0)$ matrices for Model 1 would be of the form
\begin{equation}
    K =\left(\begin{array}{cc}
      W & 0 \\
      0 & W \\
\end{array}
\right)
\end{equation}
with $W$,
\begin{equation}
    W = \left(\begin{array}{ccc}
      \omega_{a} &0  & 0   \\
     0  & \omega_{s} & g_{i} \\
     0  & g_{i} & \omega_{r_{i}}   \\
    \end{array} 
\right),
\end{equation}
and $\gamma (0)$, 
\begin{equation}
\gamma (0) = \left(\begin{array}{cc}
                  A_{+} & 0 \\
                  0 & A_{-} \\
                \end{array}
\right),
\end{equation}
where  $A_{\pm}$ is
\begin{equation}
    A_{\pm} =\left(\begin{array}{ccc}\
     \cosh{\zeta} & \pm\sinh{\zeta} &  0   \\
                  \pm\sinh{\zeta} & \cosh{\zeta} & 0    \\
                   0& 0 & 1+\frac{2}{e^{\omega_{r_{i}}/T}-1}    \\
                   \end{array}
\right).
\end{equation}
Unless stated otherwise the fixed parameters used in this paper are the system and ancilla frequencies, $\omega_{s} = \omega_{a} = 10$, the maximum bath mode frequency, $\omega_{bmax}= 50$, the squeezing parameter , $\zeta=4$, and the temperature, $T=1$. The cut-off frequencies, $\omega_{c}$, for the Ohmic and super Ohmic baths are 15 and 3 respectively.
\titlespacing*{\section}{0pt}{10pt}{10pt}
\section{Appendix B - Analytic expression for the entanglement dynamics of a coupled oscillator}
We begin by considering Eq. \eqref{genham} for one `bath' mode, i.e. $i=1$, and then we can move into the interaction picture for a coupled oscillator model. We achieve this by making a rotation of $e^{-At}$ on the wavefunction to make it time dependant. This new wavefunction, $\Psi(t)=e^{-iAt}\Psi$, is then inserted into the Schr\"{o}dinger equation,
\begin{equation}
\begin{aligned}
i\frac{\partial \Psi}{\partial t} &= (A + e^{-iAt}H e^{iAt})\Psi \\
i\frac{\partial \Psi}{\partial t} &= \widetilde{H}\Psi. \\
\end{aligned}
\end{equation}
Choosing $A=-\omega_{s}(a^{\dagger}a + s^{\dagger}s + r^{\dagger}r)$ and using the Hadamard Lemma we get the interation picture Hamiltonian
\begin{equation}
\begin{aligned}
\widetilde{H} &=  \Delta r^{\dagger}r + g(sr^{\dagger} + s^{\dagger}r), \\
\end{aligned}
\end{equation}
where we have used the fact that $\omega_{a} = \omega_{s}$ for all our models, and that the detuning $\Delta = (\omega_{r}-\omega_{s})$. 
The next step is to find the eigenvalues and eigenvectors of the Hamiltonian. To this end, we define the normal modes of the form $p_{i} = v_{i}s + w_{i}r$ where $v$ and $w$ are factors to be  determined by enforcing the commutation relation $[p_{i},p^{\dagger}_{i}] \equiv 1$. Finding the eigenvectors and eigenvalues for the normal modes, we arrive at the following set of equations expressed in matrix form
 \begin{equation}
\left( \begin{array}{c}
p_{1}  \\   
p_{2} \\
\end{array} \right)
= \left( \begin{array}{cc}
A & B  \\
C & D  \\
\end{array} \right) 
\left( \begin{array}{c}
s \\
r \\
\end{array} \right) .
\end{equation}
\begin{equation}
\begin{array}{cc}
A = \sqrt{\frac{2g^2}{E_{p}(E_{p} + \Delta)}} & B = \sqrt{\frac{E_{p} + \Delta}{2E_{p}}} \\
 C = \sqrt{\frac{2g^2}{E_{p}(E_{p} - \Delta)}} & D = -\sqrt{\frac{E_{p} - \Delta}{2E_{p}}} \\
\end{array}
\end{equation}
\begin{equation}
\begin{aligned}
E_{p} &= \sqrt{\Delta^{2} + 4g^{2}} \\
\end{aligned}
\end{equation}
To find the time dependence of the system and oscillator modes, we utilise the knowledge that the normal $p$ modes evolve in the following way;
\begin{equation}
\begin{aligned}
\frac{\partial p_{i}}{\partial t} &= -i [p_{i},H] \\
\frac{\partial p_{i}}{\partial t} &= -i E_{i}p_{i} \\
p_{i}(t) &= p_{i} e^{-iE_{i}t}. \\ 
\end{aligned}
\end{equation} 
This allows us to write the following time evolutions 
\begin{equation}
\begin{aligned}
s(t) &= s\left(\cos\left(\frac{E_{p}t}{2}\right) + \frac{i\Delta}{E_{p}}\sin\left(\frac{E_{p}t}{2}\right)\right)e^{\frac{-i\Delta t}{2}} \\
&- r\left(\frac{2ig}{E_{p}}\right)\left(\sin\left(\frac{E_{p}t}{2}\right)\right)e^{\frac{-i\Delta t}{2}}  \\
\end{aligned}
\end{equation}
\begin{equation}
\begin{aligned}
r(t)&= - s\left(\frac{2ig}{E_{p}}\right)\left(\sin\left(\frac{E_{p}t}{2}\right)\right)e^{\frac{-i\Delta t}{2}} \\
&+ r\left(\cos\left(\frac{E_{p}t}{2}\right) - \frac{i\Delta}{E_{p}}\sin\left(\frac{E_{p}t}{2}\right)\right)e^{\frac{-i\Delta t}{2}} \\
\end{aligned}
\end{equation}
If we have an initial state of the form shown in Appendix A with $i=1$, using the time evolution of the above modes we can construct a matrix which acts on the intial covariance matrix. This will give us the time evolution of the covariance matrix which tells us how the state of the model is changing with time. Below is the time dependant covariance matrix for the ancilla-system state, where we have left out the martix components of the bath mode.
\begin{equation}
\gamma (t) = \frac{1}{2}\left(\begin{array}{cc}
\it{V_{1+}} & \it{V_{2}} \\
\it{V_{2}} & \it{V_{1-}}\\
\end{array}
\right)
\end{equation}
\begin{equation}
\it{V_{1\pm}} = \left(\begin{array}{cc}
\cosh\left(\zeta\right) & \pm \Xi \sinh \left(\zeta \right) \\
\pm \Xi  \sinh \left(\zeta \right) & \cosh\left(\zeta\right) +\Phi \\
 \end{array}
\right)
\end{equation}
\begin{equation}
\it{V_{2}} = \left(\begin{array}{cc}
0 & \Pi \sinh \left(\zeta \right) \\
 \Pi \sinh \left(\zeta \right) & 0 \\
\end{array}
\right)
\end{equation}
\begin{equation}
\begin{aligned}
\Xi &= \left(\cos\left(\frac{E_{p}t}{2}\right)\cos\left(\Delta t\right)  + \frac{ \sin\left(\frac{E_{p}t}{2}\right)\sin\left(\Delta t\right)\Delta}{E_{p}}\right)\\
\Pi &= \left(-\cos\left(\frac{E_{p}t}{2}\right)\sin\left(\Delta t\right)  + \frac{ \sin\left(\frac{E_{p}t}{2}\right)\cos\left(\Delta t\right)\Delta}{E_{p}}\right) \\
\Phi &=  \frac{2g^{2}\left(\cos\left(E_{p}t\right) - 1\right)}{E^{2}_{p}}\left(\cosh\left(\zeta\right) - \coth\left(\frac{\omega_{r}}{2}\right)\right) \\
\end{aligned}
\end{equation}
As shown in Ref. \cite{covareq},  from this matrix we can calculate how the system-ancilla entanglement is evolving by finding the symplectic eigenvalues (from which we can calculate the logarithmic negativity). Then by using a second order expansion of $g/\Delta$ in the formula for calculating the symplectic eigenvalues we can find an analytic form of the entanglement dynamics
\begin{equation}
\mathbb{E}= \log_{2}\left(\frac{\Delta^{2}e^{-\zeta}+ 2g^{2}\left(\coth\left(\frac{w_{r}}{2}\right) - e^{-\zeta} \right)\sin\left(\frac{\Delta t}{2}\right)}{2\Delta^{2}}\right).
\end{equation}
\titlespacing*{\section}{0pt}{10pt}{10pt}
\section{Appendix C - Effective Hamiltonians for the two 'bath' modes case}

For the case of two `bath' modes, equivalent Hamiltonians for small, large and zero detuning cases can be found, which should give us an insight into the behaviour. Considering the case where we have one resonant ($b$) and one off-resonant ($r$) bath mode the Hamiltonian is
\begin{equation}
H = \Delta r^{\dagger}r + g(sb^{\dagger} + s^{\dagger}b) + h(sr^{\dagger} + s^{\dagger}r).
\end{equation}
For the large detuning case, performing a rotation by $\Delta r^{\dagger}r$ the interaction Hamiltonian can be expressed as
\begin{equation}
\widetilde{H} = g(sb^{\dagger} + s^{\dagger}b) + h(s^{\dagger}re^{-i\Delta t} + sr^{\dagger}e^{i\Delta t}).
\end{equation}
We then utilise an approximation, outlined in a paper by Gamel and James \cite{James}, which averages over the fast oscillating exponential terms to give the effective Hamiltonian
\begin{equation}
H_{eff} = \frac{h^{2}}{\Delta}s^{\dagger}s + (\Delta -\frac{h^{2}}{\Delta}) r^{\dagger}r + g(sb^{\dagger} + s^{\dagger}b).
\end{equation}
This shows that the large detuned mode effectively decouples from the system and introduces a frequency shift in the system. Therefore the entanglement dynamics can be modelled with just one `bath' mode with a small detuning. 

The procedure can be repeated for the case of small detuning, where we now rotate by $g(sb^{\dagger} + s^{\dagger}b) + h(sr^{\dagger} + s^{\dagger}r)$ and average over all exponentials of the form $\exp[\pm i t \sqrt{g^{2}+h^{2}}]$. This leads to the Hamiltonian
\begin{equation}
\begin{aligned}
H_{eff} &= \left(\frac{\Delta h^{2}}{2(g^{2}+h^{2})}\right)s^{\dagger}s + \left(\frac{3\Delta g^{2}h^{2}}{2(g^{2}+h^{2})^{2}}\right)b^{\dagger}b \\
&+ \left(\frac{\Delta(2g^{4}+h^{4})}{2(g^{2}+h^{2})^{2}}\right) r^{\dagger}r + g(sb^{\dagger} + s^{\dagger}b) + h(sr^{\dagger} + s^{\dagger}r) \\
&+ \left(\frac{\Delta gh(h^{2}-2g^{2})}{2(g^{2}+ h^{2})^{2}}\right)(br^{\dagger} + b^{\dagger}r), \\
\end{aligned}
\end{equation} 
suggesting that the we get frequency shifts in all the modes and an effective coupling is gained between the two `bath' modes. This is understandable as the communication speeds to both `bath' modes from the system would be similar (if the couplings are of the same order) due to the small detuning, so that an effective coupling is created. Note that these new features of the Hamiltonian would mean that rather complicated entanglement dynamics is at play, because the `bath' has the ability to hold onto the entanglement over time. There were various attempt to find an exact analytic solution for the two `bath' mode case, but this proved to be too complex due to the lengthy forms of the eigenvalues and eigenvectors. 

In the case where the system is coupled to only to resonant modes, using a normal mode transformation we get the Hamiltonian
\begin{equation}
H_{eff}=\sqrt{g^{2}+h^{2}}\left(sp^{\dagger}+s^{\dagger}p\right).
\end{equation}
That is, one of the normal modes couples to the system and other decouples and is now a dark mode which only undergoes free evolution. Therefore the model can be reduced to a one `bath' mode model with a new coupling strength. This is true regardless of the number of coupled resonant modes, where the coupling strength would be given by
\begin{equation}
g'^{2}= \sum_{i}^{N} g_{i}^{2},
\end{equation}
where N is the number of resonant bath modes. In this case all but one of the normal modes would have effectively decoupled from the system.

In terms of the effect this has on the NMBQ, in the resonant mode case we would get an increase in NMB the more resonant modes we attach to the system, due to the increased effective coupling strength. In the case of large detuning it would be similar to the NMB observed in the one resonant `bath' mode case (due to the small effective detuning). For small detuning we can see that the NMB will depend on the relative coupling strengths between, effectively, the three off-resonant modes.
\titlespacing*{\section}{0pt}{10pt}{10pt}
\section{Appendix D - Figures showing the bath mode occupancy }

\begin{figure}[!ht]
\centering
\includegraphics[width=0.45\textwidth, keepaspectratio]{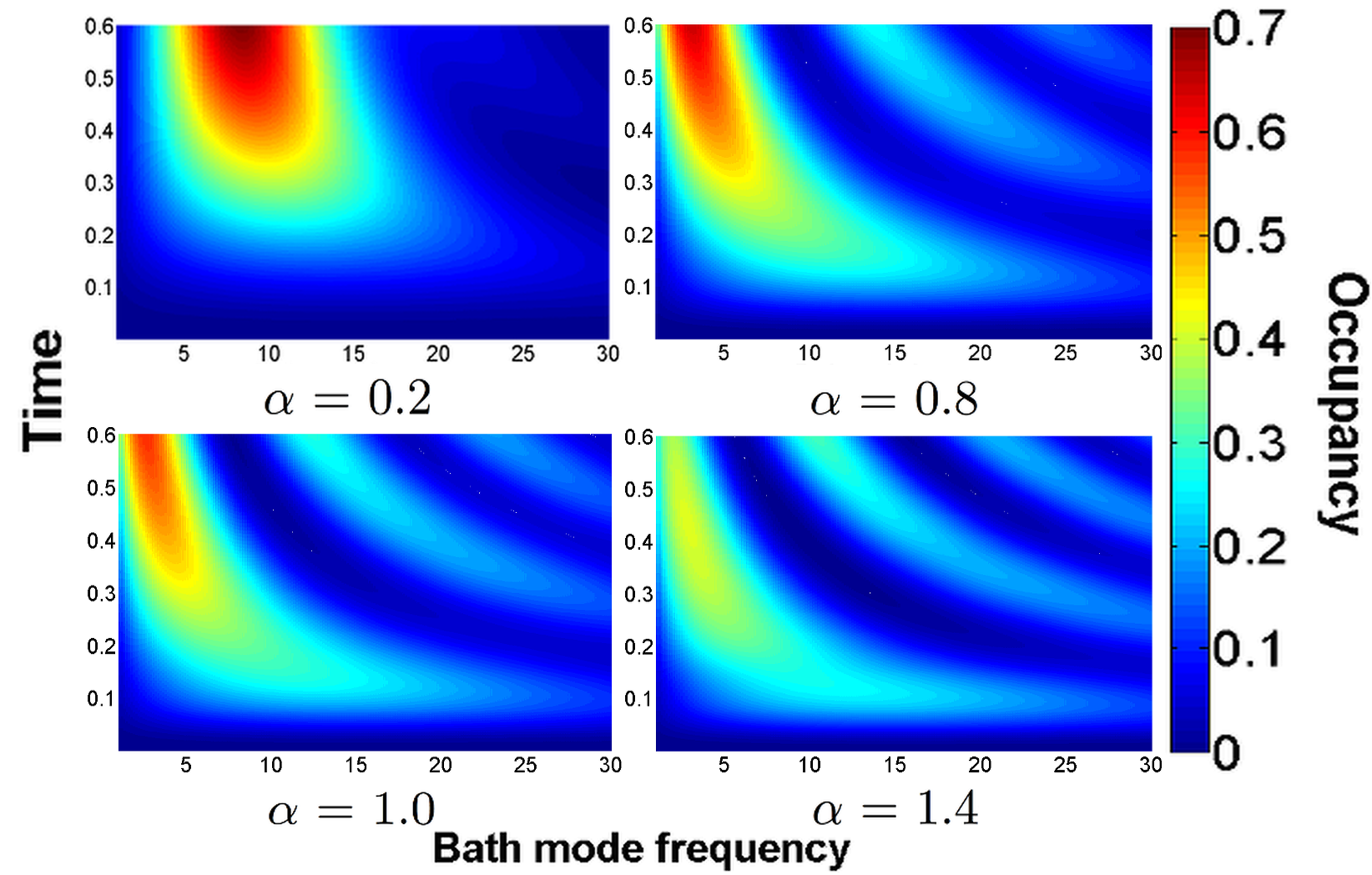}
\caption{Model 1 - Ohmic bath}\label{Mod1occ}
\end{figure}

\begin{figure}[!ht]
\centering
\includegraphics[width=0.45\textwidth, keepaspectratio]{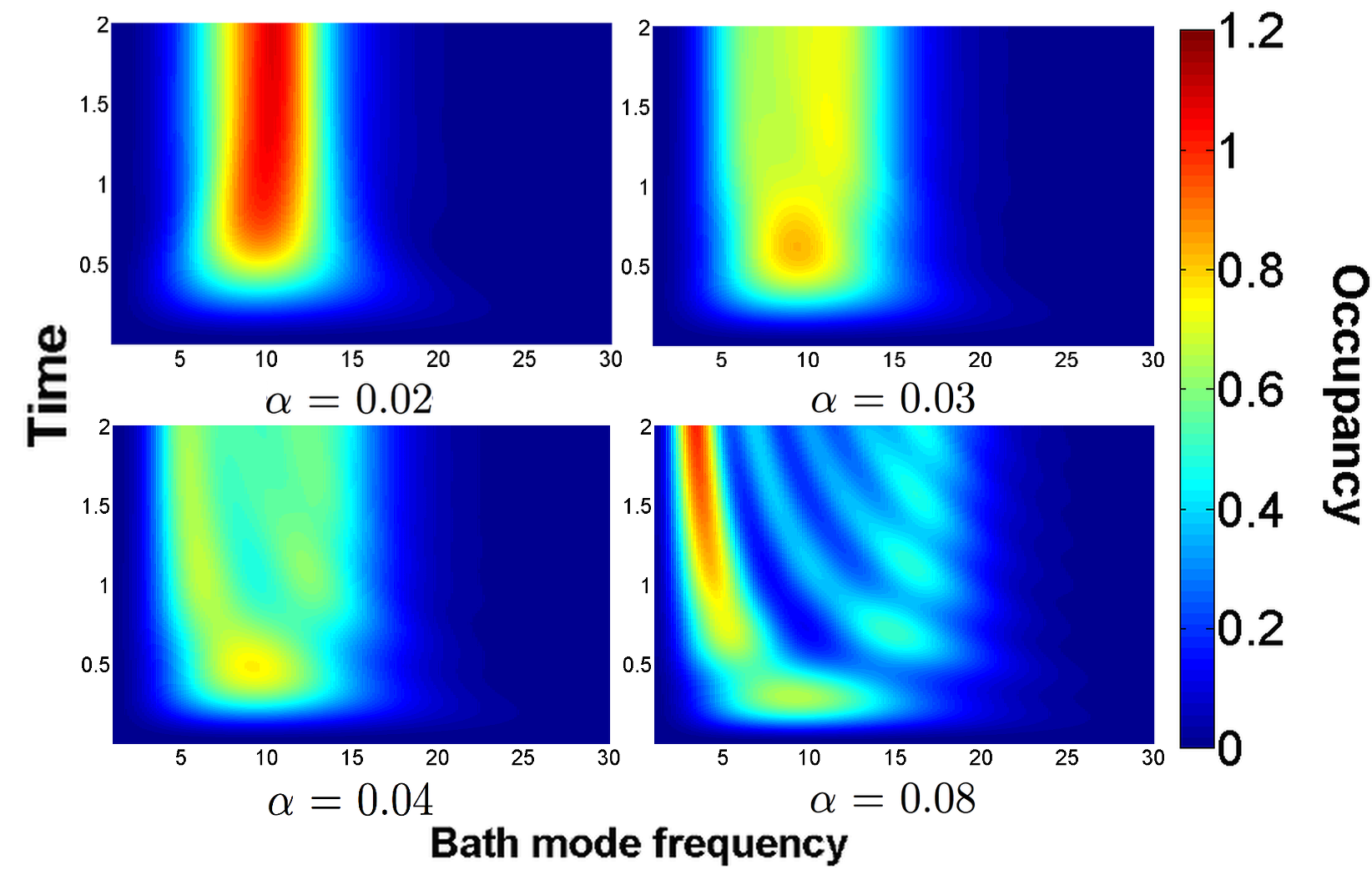}
\caption{Model 1 - Super Ohmic bath}\label{Mod1SOocc}
\end{figure}

\begin{figure}[!ht]
\centering
\includegraphics[width=0.45\textwidth, keepaspectratio]{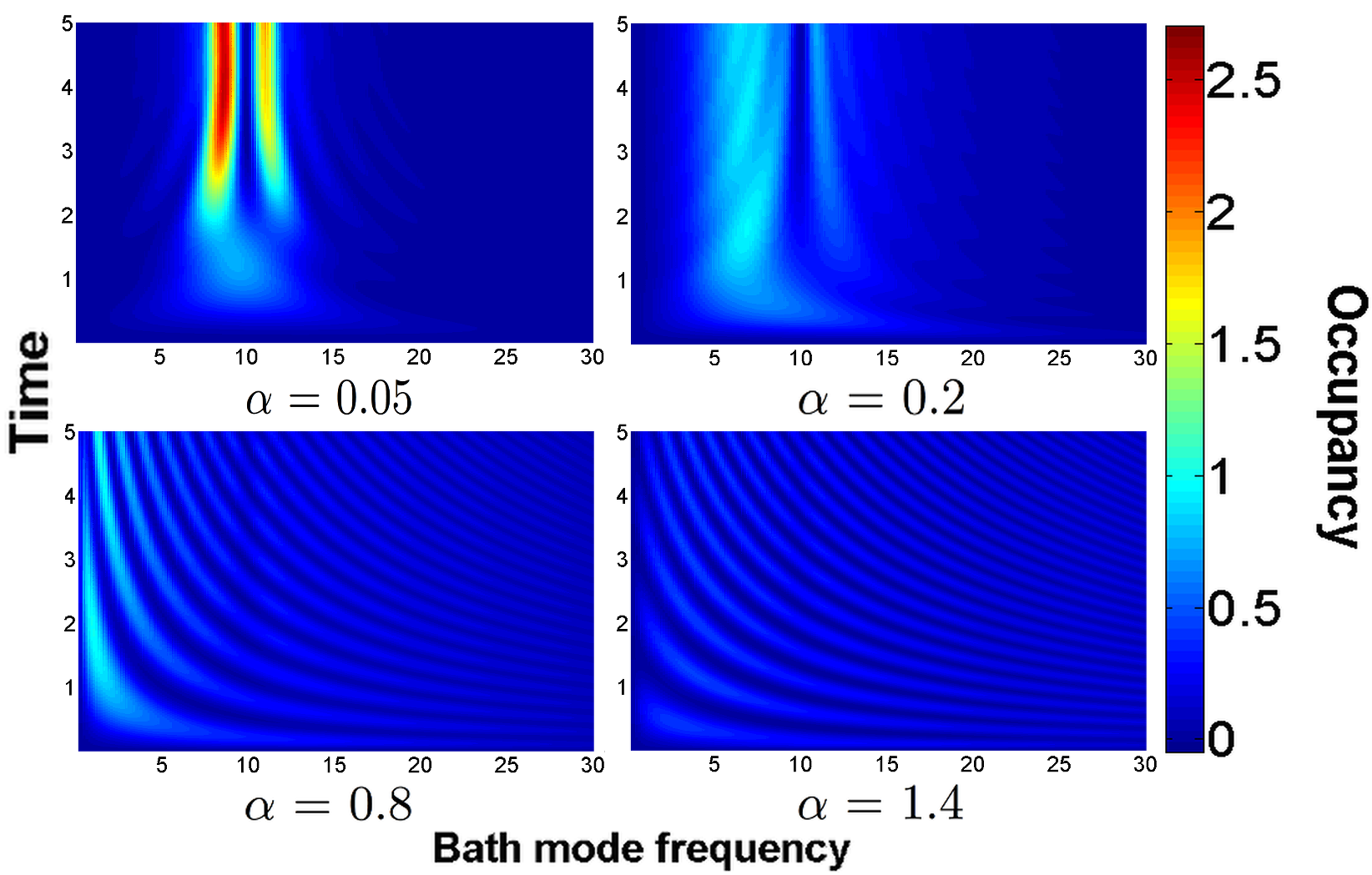}
\caption{Model 2}\label{Mod2occ}
\end{figure}

\begin{figure}[H]
\centering
\includegraphics[width=0.45\textwidth, keepaspectratio]{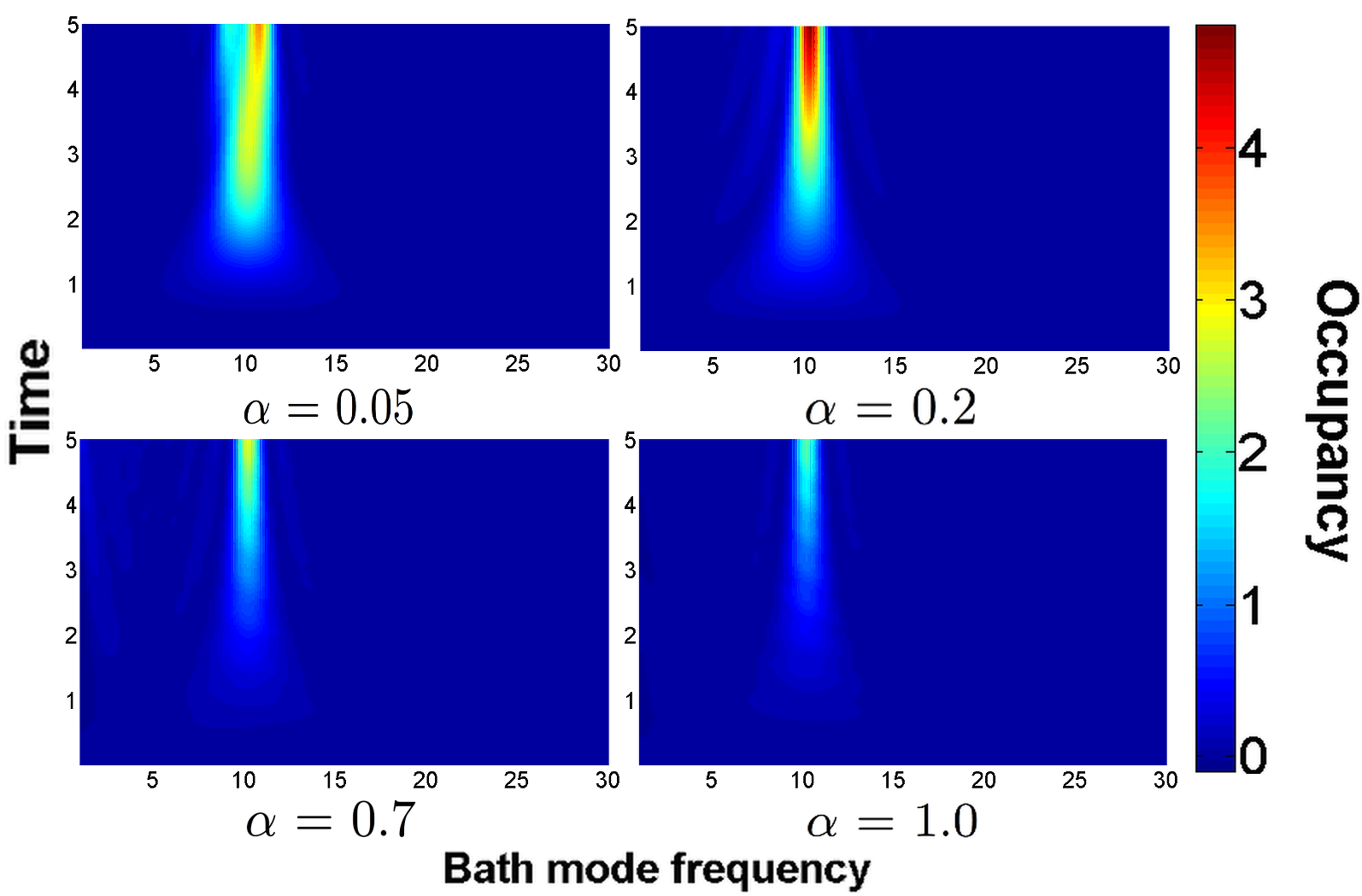}
\caption{Model 3}\label{Mod3occ}
\end{figure}

The following plots shows how the occupancy of the bath modes vary with time. The occupancy can be calculated from the covariance matrix by taking the $\hat{x}^{2}_{r_{i}}$ and $\hat{p}^{2}_{r_{i}}$ components of the matrix. In our plots we have taken the $\hat{x}^{2}_{r_{i}}+ \hat{p}^{2}_{r_{i}}$ values of the bath mode $r_{i}$ minus its initial thermal energy (i.e. the same value at $t_{0}$) as the occupancy to improve the clarity of the colour scale. It shows how much energy (up to a factor) a mode $r_{i}$ has gained from the system and therefore is an indicator of the level of interaction between the system and bath mode. Figures \ref{Mod1occ}, \ref{Mod1SOocc}, and \ref{Mod3occ} have plot ranges of 1 to 30 for $\omega_{r}$, the first few modes are ignored because they have a high initial thermal energy and so gain a negative value over time which skews the colour map on the surface plot. The high end is ignored because no significant dynamics take place in that region. Figure \ref{Mod2occ} includes all the modes in the low end because the dynamics are important in that region.
\newpage
\titlespacing*{\section}{0pt}{10pt}{10pt}
\section{Appendix E - Fidelity measure}

Another quantifier we can consider is the fidelity based measure of NMB, proposed by Vasile et. al. \cite{breuer2} as an extension of Breuer's measure \cite{breuer1} for Gaussian states. Based on the information back-flow definition of NMB, this measure has the advantage of providing a necessary and sufficient condition for NMB (hence it is a proper {\it measure}), however it relies on a maximisation step which makes it hard to compute. The measure is based on the distinguishability of two different initial states $\rho_{1}$ and $\rho_{2}$ under the action of a dynamical map $\mathcal{E}$. Under the action of any CP map, the distinguishability,   $\mathcal{D}(\rho_{1},\rho_{2})\equiv1-{\cal  F}(\rho_1,\rho_2)$, with ${\cal F}(\rho_1,\rho_2)\equiv{\sf Tr}\sqrt{\sqrt{\rho_1}\rho_2\sqrt{\rho_1}}$ the fidelity, follows the contractive property
\begin{equation}
\mathcal{D}(\mathcal{E}\rho_{1},\mathcal{E}\rho_{2})\leq\mathcal{D}(\rho_{1},\rho_{2}).
\end{equation}
Hence, under Markovian behaviour the divisibility property of Eq.~\eqref{complaw} will ensure a monotonic decrease of distinguishability, in analogy to what we observed for the system-ancilla entanglement. Such irreversible loss of distinguishability may be understood as the leakage of  `quantum information' into the bath, which in turn is unable to transfer it back to the system. Hence, any increase in distinguishability can be interpreted as the environment returning part of the leaked information to the system, a signature of NMB. Similarly to the NMBQ, a measure of non-Markovianity can be constructed by summing the distinguishability increases between pairs of quantum states. Restricting the analysis to Gaussian states, the non-Markovianity is given by;
\begin{equation}\label{fidmeasure}
\mathcal{N}_{P}= \max_{P}(-\int_{\dot{\mathcal{f}}<0} \frac{d}{dt}\mathcal{F}(P,t) dt),
\end{equation}
where we maximise over all parameters, $P$, of the Gaussian state. These parameters are not bounded and therefore running this measure  can become numerically challenging. In their paper, Vasile et. al. use a weak-coupling, secular non-Markovian master equation (\cite{HuPazZhang, Sabrina4, Sabrina2, Sabrina3, BandPbook}) to describe a model of a one mode squeezed state coupled to a bath of oscillators, but again to avoid using these approximations we utilise a large finite bath to simulate this model via a covariance matrix approach. 

\begin{center}
\begin{figure}[h!]
\centering
\includegraphics[width=\textwidth,height=6.3cm,keepaspectratio]{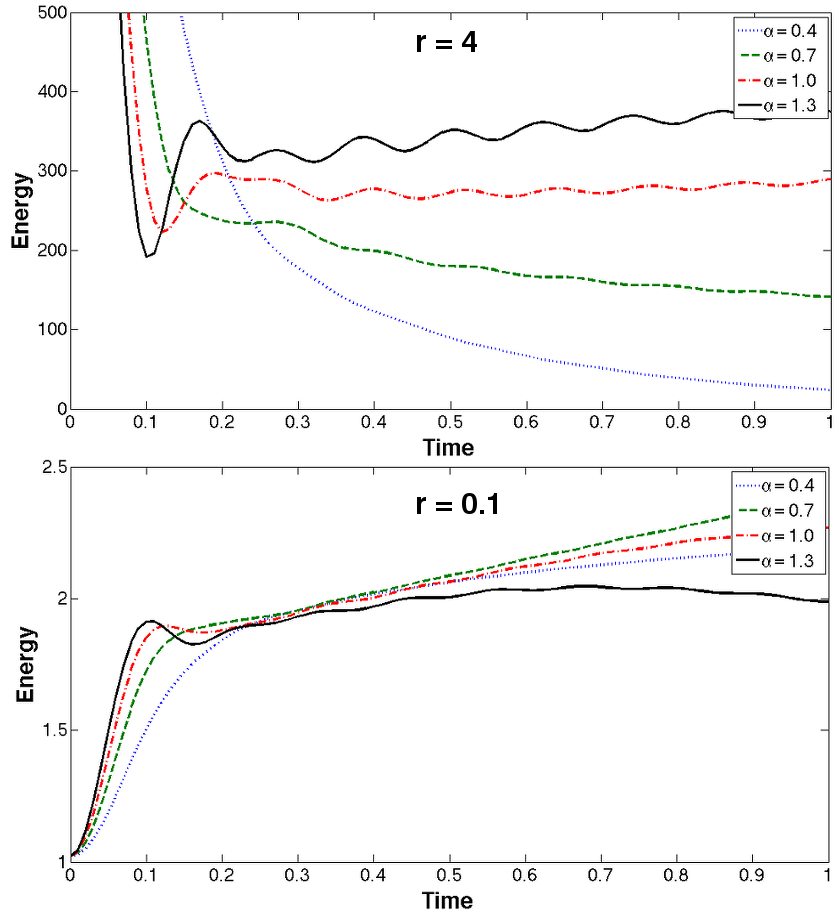}
\caption{Energy dynamics of one mode squeezed states coupled to an Ohmic bath. The initial squeezing parameter of the states are 4 and 0.1. We find oscillations in the energy dynamics in a similar fashion to the EO, in the NMBQ case, as $\alpha$ is varied.} \label{energyr4and0_1}
\end{figure}
\end{center}

\begin{center}
\begin{figure}[h!]
\centering
\includegraphics[width=0.5\textwidth]{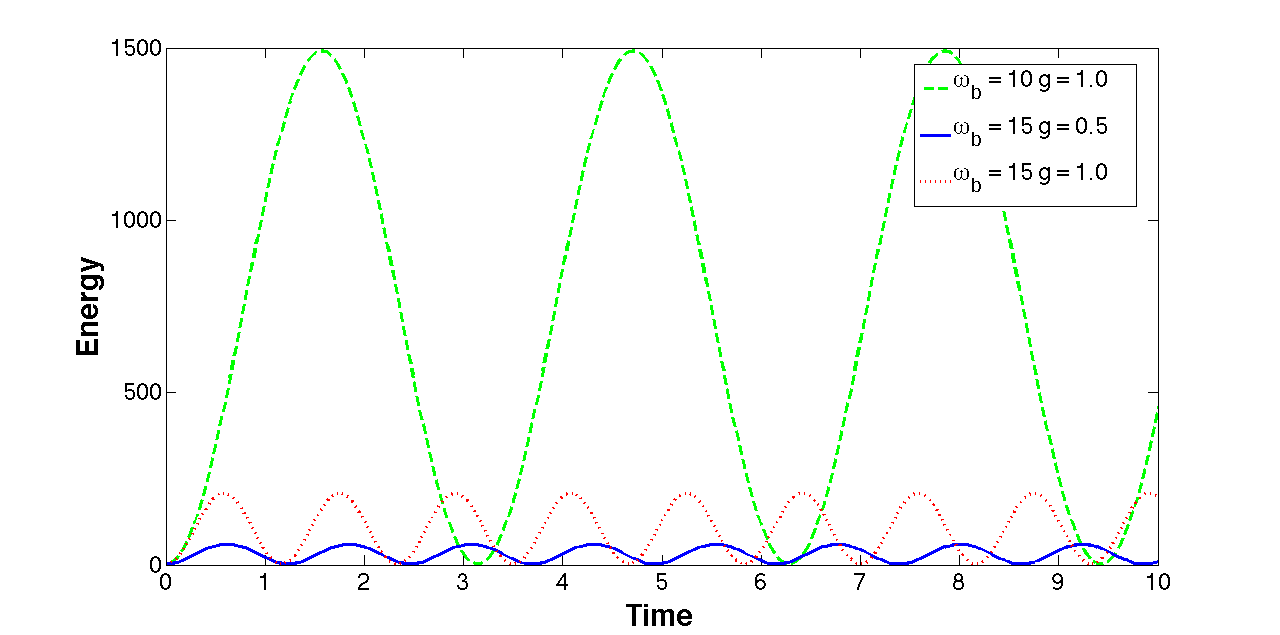}
\caption{Energy dynamics of one mode squeezed state (with squeezing parameter 4) coupled to a thermal mode with temperature $T =1$. The three lines represent the energy dynamics of the system mode; Green - [$\omega_{b}= 10$ , $g = 1$], Blue - [$\omega_{b}= 15$ , $g = 0.5$], Red - [$\omega_{b}= 15$ , $g = 1$.]. } \label{2modengdynm}
\end{figure}
\end{center}

The fidelity relies on the states of the two systems and is therefore dependant on the energy of the states at any given time. Since the measure collec
ts variations in the fidelity, the energy dynamics could have an impact on the NMB. We find that the energy dynamics of the system is dependant on the the modes in a similar fashion to that of entanglement. But these dynamics are very dependant on the initial state of the system, which is why the maximisation procedure is needed. Even if we restrict ourselves to squeezed states, a maximisation over both the squeezing parameter and the phase is necessary, which makes the measure numerically time consuming. What can be done, is to understand the impact the modes would have on the predictions of this measure. Note that under the divisibility definition this measure, even with the maximisation, would only be a witness, as there exisits non-divisible dynamics which can increase the fidelity. 
\begin{center}
\begin{figure}[h!]
\centering
\includegraphics[width=0.45\textwidth]{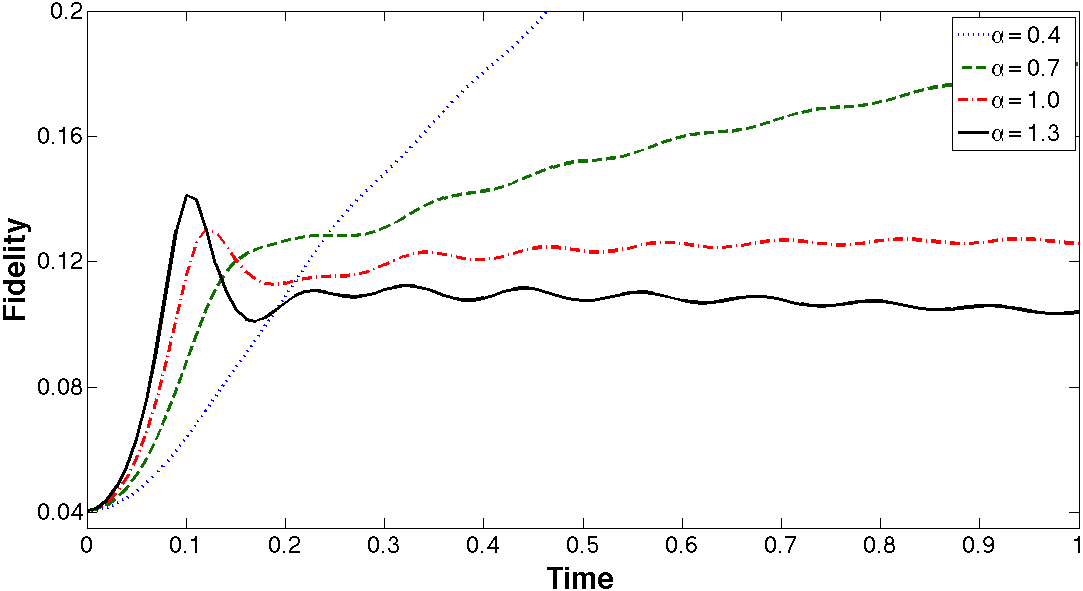}
\caption{Fidelity dynamics between two different initial one mode squeezed states. The two states have zero phase but they have different squeezing parameters, r, of 4 and 0.1. We find more oscillations in the fidelity as $\alpha$ is increased.} \label{fidr4and0_1}
\end{figure}
\end{center}
If we just invesigate one pair of inital states, the fidelity can still be used as a quantifier of NMB for this particular case under their definition of NMB. We choose the system mode in a single mode squeezed state with two different squeezing parameters, r,  and the bath in an thermal state with temperature, $T=1$. The spectral density is Ohmic and all associated frequency parameters are the same as the models in the main text with the frequency of the system mode kept at 10. Figure \ref{fidr4and0_1} shows the fidelity dynamics for a pair of intial states which vary only in r, with values of 4 and 0.1.  The oscillations in the fidelity seem to coincide with the oscillations in the energy dynamics (Figure \ref{energyr4and0_1}), indicating that the energy dynamics do have a role to play in the NMB predictied by the fidelity quantifier/measure. 
Figure \ref{energyr4and0_1} also shows that as the $\alpha$ value increases more oscillations are seen in the energy of the system. Taking the $r=4$ case, we find that in similar fashion to the entanglement dynamics, the system shares energy with the same dependence on coupling strength and frequency of the system as in the two mode case (Figure \ref{2modengdynm}). In the many mode case, for $r=4$, we find that at low couplings the near resonant modes get the majority of the energy and therefore affect the NMB, but as coupling is increased the detuned modes start to gain more energy from the system (which can be seen from the occupancy figures in Appendix D) and become the driving force of the NMB due to the high frequency of their energy oscillations.  
\newpage

\end{document}